\documentclass[twocolumn]{aastex63}

\newcommand{\msun}{M$_{\sun}$}

\usepackage{graphicx}
\usepackage[caption=false]{subfig}
\usepackage{xcolor}
\usepackage{url}
\usepackage{hyperref}
\usepackage{breqn}

\newcounter{column_number}
\shortauthors{Xin et al.}
\shorttitle{Dust Masses of Protoplanetary Disks in Lupus}

\begin{document}

\title{Measuring the Dust Masses of Protoplanetary Disks in Lupus with ALMA: Evidence that Disks can be Optically Thick at 3 mm}

\correspondingauthor{Zihua Xin}
\email{zihuaxin@bu.edu}

\author[0000-0001-6060-2730]{Z. Xin}
\affil{Institute for Astrophysical Research, Department of Astronomy, Boston University, 725 Commonwealth Avenue, Boston, MA 02215, USA}

\author[0000-0001-9227-5949]{C. C. Espaillat}
\affil{Institute for Astrophysical Research, Department of Astronomy, Boston University, 725 Commonwealth Avenue, Boston, MA 02215, USA}

\author[0000-0002-3091-8061]{A. M. Rilinger}
\affil{Institute for Astrophysical Research, Department of Astronomy, Boston University, 725 Commonwealth Avenue, Boston, MA 02215, USA}

\author[0000-0003-3133-3580]{{\'A}. Ribas}
\affil{Institute of Astronomy, University of Cambridge, Madingley Road, Cambridge CB3 0HA, UK}
\affil{European Southern Observatory (ESO), Alonso de Córdova 3107, Vitacura, Casilla, 19001, Santiago de Chile, Chile}

\author[0000-0003-1283-6262]{E. Mac{\'i}as}
\affil{European Southern Observatory (ESO), Karl-Schwarzschild-Straße 2, 85748 Garching bei München, Germany}

\begin{abstract}

Accurate disk mass measurements are necessary to constrain disk evolution and the timescale of planet formation, but such measurements are difficult to make and are very dependent on assumptions. Here we look at the assumption that the disk is optically thin at radio wavelengths and the effect of this assumption on measurements of disk dust mass.
We model the optical to radio spectral energy distributions (SEDs) of 41 protoplanetary disks located in the young ($\sim1$--3 Myr old) Lupus star-forming region, including 0.89~mm, 1.33~mm, and 3~mm flux densities when available.  We measure disk dust masses that are $\sim1.5$--6 times higher than when using the commonly adopted disk dust mass equation under the assumption of optically thin emission in the (sub-)millimeter. The cause of this discrepancy is that most disks are optically thick at millimeter wavelengths, even up to 3~mm, demonstrating that observations at longer wavelengths are needed to trace the fully optically thin emission of disks.

\end{abstract}

\keywords{accretion disks, stars: circumstellar matter, 
planetary systems: protoplanetary disks, 
stars: formation, 
stars: pre-main sequence}

\section{Introduction}

Protoplanetary disks contain the reservoirs of gas and dust necessary to form planets, providing important insights on the conditions for planet formation. In particular, the amount of material in the disk and the length of time this material is present set an upper limit on the timescale for planet formation. Studies of nearby star-forming regions with the {\it Spitzer Space Telescope} measured infrared (IR) emission from small micron-sized grains in the upper atmosphere of the inner disk and found that protoplanetary disks disperse within 5--10 Myr \citep[e.g.,][]{haisch01b, hernandez07a, ribas15}, establishing an upper limit to the timescale for giant planet formation.  More recently, ALMA has given us the ability to measure the amount of millimeter-sized grains in protoplanetary disks using millimeter flux densities on a large scale. Recent ALMA surveys find a decline in disk dust masses with age within a few Myr \citep{ansdell2016, pascucci16, ansdell17, long17, barenfeld16, miotello17, manara22}, seemingly in parallel with the IR studies showing a trend of decline of disk fraction with age.  

ALMA protoplanetary disk surveys find typical disk gas masses of $\sim$1 M$_{\rm Jup}$ and dust masses of 0.2--0.4 M$_{\earth}$, which are too low to support ongoing planet formation even at 1 Myr \citep[e.g.,][]{ansdell2016, pascucci16}.  This has led some to suggest that planet formation is well underway by the Class II phase \citep{najita14, manara18, mulders21}, which may also be supported by the detection of substructure in $\sim1$ Myr old disks that can be attributed to planets \citep[e.g.,][]{alma15, andrews18,seguracox20}. If planets form earlier than the Class II phase, the expectation is that  significantly more disk material is present in the earlier phases. There is indeed a reported decrease in disk dust mass from the Class I to Class II phases \citep[e.g.,][]{sheehan17, tychoniec18, tychoniec20}, 
although this decrease has recently been shown to be less clear when comparing disk dust masses from different studies that use radiative transfer modeling \citep{sheehan22}.  

To make further progress on constraining the timescale of planet formation, we need accurate measurements of the amount of disk material present. Gas dominates the mass of the disk but is very difficult to measure. A large spread in reported disk masses using gas measurements has been attributed to gas freezing out onto grains and to assumptions on chemical abundances \citep[see][for a review]{miotello22}. Dust grains are easier to observe than the gas since the dust dominates the opacity of the disk.  However, dust masses are also hard to constrain, and differences in reported masses have been attributed to different assumed temperatures and dust opacities. In addition, it has recently been seen that disks can be optically thick at millimeter wavelengths \citep[e.g.,][]{tripathi17, macias21}, and most measurements of disk dust mass are made using millimeter flux densities along with the assumption that the disk is optically thin at millimeter wavelengths.

Here we measure disk dust masses in the Lupus star-forming region using spectral energy distribution (SED) modeling, incorporating optical to radio fluxes and {\it Spitzer} IR spectra when available. Lupus is a young (1--3~Myr), low-mass star-forming region \citep{comeron2008,alcala17} at a median distance of 158.5~pc \citep{gaia18}. It is composed of four main star-forming clouds (Lupus I--IV) and has been well studied in the IR and radio. The c2d {\it Spitzer} legacy project \citep{evans2009} observed Lupus I, II, and IV, and ALMA observed Lupus I--IV \citet{ansdell2016}.  We follow \citet{ribas2020} and employ the D'Alessio irradiated accretion disk \citep[DIAD, ][]{dalessio98, dalessio99, dalessio01, dalessio05, dalessio06} models with an artificial neural network (ANN). In Section~2 we explain how we selected our sample.  In Section~3 we review the DIAD models and the ANN utilized in this work. We present our results in Section~4 and discuss these results in Section~5. We conclude with a summary in Section~6.

\begin{deluxetable*}{ccc}[ht!]    
\tablecaption{Sample Photometry \label{tab:photo}}
\tablehead{
\colhead{Wavelength} & \colhead{Instrument} & \colhead{Reference} \\
\colhead{$(\mu m)$}
}
\startdata
0.623 & {\it Gaia} DR2 & \citet{gaia18} \\
0.625 & AAVSO & \citet{aavso} \\
0.764 & AAVSO & \citet{aavso} \\
0.773 & {\it Gaia} DR2 & \citet{gaia18} \\
1.25 & 2MASS & \citet{2mass} \\
1.65 & 2MASS & \citet{2mass} \\
2.17 & 2MASS & \citet{2mass} \\
3.35 & {\it WISE} & \citet{wise}\\
3.6 & {\it IRAC} & \citet{c2d}\\
4.5 & {\it IRAC} & \citet{c2d}\\
4.6 & {\it WISE} & \citet{wise}\\
5.8 & {\it IRAC} & \citet{c2d}\\
8.0 & {\it IRAC} & \citet{c2d}\\
11.6 & {\it WISE} & \citet{wise}\\
22.1 & {\it WISE} & \citet{wise}\\
24 & {\it MIPS} & \citet{c2d}\\
70 & {\it MIPS} & \citet{c2d}\\
160 & {\it Herschel} & \citet{Benedettini2018} \\
250 & {\it Herschel} & \citet{Benedettini2018} \\
350 & {\it Herschel} & \citet{Benedettini2018} \\
500 & {\it Herschel} & \citet{Benedettini2018} \\
890 & ALMA & \citet{ansdell2016} \\
1333 & ALMA & \citet{ansdell2018} \\
3000 & ALMA & \citet{tazzari2021}
\enddata
\end{deluxetable*}

\section{Sample}

\begin{deluxetable*}{cccccccc}    
\tablecaption{Stellar Parameters \label{tab:stellarparam}}
\tablehead{
\colhead{Object} & \colhead{SpT} & \colhead{$T_{\mathrm{eff}}$} & \colhead{$A_{\mathrm{v}}$} & \colhead{Mass} & \colhead{Radius} & \colhead{Parallax} & \colhead{Inclination}\\
\colhead{} & \colhead{} & \colhead{(K)} & \colhead{} & \colhead{$(M_\odot)$} & \colhead{$(R_\odot)$} & \colhead{(mas)} & \colhead{(deg)}}
\startdata
J15450634-3417378   &--             &--             &2.9    &--             &--             &--&43$\pm$28\\
J15450887-3417333   &M5.5$\pm$0.5   &3060$\pm$71    &5.5    &0.14$\pm$0.03  &0.9$\pm$0.2    &6.5$\pm$0.1&45$\pm$4\\
J15560210-3655282   &-- &-- &-- &0.46$\pm$0.12  &-- &6.33$\pm$0.02  &--\\
J15592523-4235066   &M5$\pm$0.5     &3125$\pm$72    &0      &0.12$\pm$0.03  &0.5$\pm$0.1  &6.79$\pm$0.07&--\\
J16000060-4221567   &M4.5$\pm$0.5   &3200$\pm$74    &0      &0.19$\pm$0.03  &1.0$\pm$0.2  &6.27$\pm$0.03&--\\
J16000236-4222145   &M4$\pm$0.5     &3270$\pm$75    &4.6    &0.24$\pm$0.03  &1.2$\pm$0.3  &6.23$\pm$0.04&65$\pm$0\\
J16002612-4153553   &M5.5$\pm$0.5   &3060$\pm$71    &0.9    &0.14$\pm$0.03  &0.9$\pm$0.3  &6.13$\pm$0.05&--\\
J16030548-4018254   &M0$\pm$0.5	&3850$\pm$177	&1.1	&0.56$\pm$0.14	&2.49$\pm$0.58	&6.46$\pm$0.02&--\\
J16070384-3911113   &M5.5$\pm$0.5   &3200$\pm$74    &0.6    &0.10$\pm$0.03  &0.23$\pm$0.06  &--&--\\
J16073773-3921388   &M5.5$\pm$0.5   &3057$\pm$70    &0      &0.11$\pm$0.02  &0.5$\pm$0.1  &6.2$\pm$0.1&--\\
J16075475-3915446   &--             &--             &2.9    &--             &--             &--&--\\
J16080017-3902595   &M5.5$\pm$0.5   &3057$\pm$70    &0      &0.13$\pm$0.02  &0.8$\pm$0.2  &6.21$\pm$0.06&--\\
J16084940-3905393   &M4$\pm$0.5     &3270$\pm$75    &2.2    &0.26$\pm$0.03  &1.6$\pm$0.4  &6.24$\pm$0.10&--\\
J16085324-3914401   &M3$\pm$0.5     &3415$\pm$79    &1.9    &0.32$\pm$0.04  &1.6$\pm$0.4  &6.13$\pm$0.05&0$\pm$0\\
J16085373-3914367   &M5.5$\pm$0.5   &3060$\pm$71    &4      &0.10$\pm$0.02  &0.29$\pm$0.06  &7$\pm$2&--\\
J16085529-3848481   &M6.5$\pm$0.5   &2935$\pm$68    &0      &0.10$\pm$0.02  &1.1$\pm$0.3  &6.43$\pm$0.09&--\\
J16092697-3836269   &M4.5$\pm$0.5   &3200$\pm$74    &2.2    &0.20$\pm$0.03  &1.1$\pm$0.2  &6.28$\pm$0.07&--\\
J16124373-3815031   &M1$\pm$0.5     &3705$\pm$171   &1      &0.5$\pm$0.1    &1.9$\pm$0.4  &6.26$\pm$0.02&46$\pm$6\\
Sz 65               &K7$\pm$0.5     &4060$\pm$187   &0.6    &0.8$\pm$0.2    &1.8$\pm$0.4  &6.52$\pm$0.03&0$\pm$0\\
Sz 66               &M3$\pm$0.5     &3415$\pm$79    &1      &0.31$\pm$0.04  &1.3$\pm$0.3  &6.41$\pm$0.03&--\\
Sz 69               &M4.5$\pm$0.5   &3197$\pm$74    &0      &0.19$\pm$0.03  &1.0$\pm$0.2  &6.55$\pm$0.07&69$\pm$21\\
Sz 71               &M1.5$\pm$0.5   &3632$\pm$167   &0.5    &0.4$\pm$0.1  &1.4$\pm$0.3  &6.44$\pm$0.02&40$\pm$0\\
Sz 72               &M2$\pm$0.5     &3560$\pm$164   &0.75   &0.38$\pm$0.09  &1.3$\pm$0.3  &6.38$\pm$0.02&53$\pm$19\\
Sz 73               &K7$\pm$0.5     &4060$\pm$187   &3.5    &0.8$\pm$0.2  &1.4$\pm$0.3  &6.34$\pm$0.03&48$\pm$3\\
Sz 75   &K6$\pm$0.5	&4205$\pm$193	&0.7	&0.51$\pm$0.14	&2.26$\pm$0.53	&6.49$\pm$0.03  &--\\
Sz 82               &M0$\pm$0.5     &3060$\pm$71    &--     &0.6$\pm$0.1  &2.7$\pm$0.7  &6.42$\pm$0.02&--\\
Sz 83               &K7$\pm$0.5     &4060$\pm$200   &0      &0.8$\pm$0.2  &2.4$\pm$0.6  &6.35$\pm$0.04&0$\pm$0\\
Sz 90               &K7$\pm$0.5     &3900$\pm$187   &3.2    &0.8$\pm$0.2  &1.6$\pm$0.4  &6.24$\pm$0.02&0$\pm$0\\
Sz 95               &M3$\pm$0.5     &3400$\pm$79    &0.7    &0.33$\pm$0.03  &1.8$\pm$0.4  &6.32$\pm$0.08&--\\
Sz 96               &M1$\pm$0.5     &4000$\pm$171   &2.4    &0.5$\pm$0.1  &2.0$\pm$0.5  &6.41$\pm$0.02&--\\
Sz 97               &M4$\pm$0.5     &3270$\pm$75    &0      &0.25$\pm$0.03  &1.3$\pm$0.3  &6.36$\pm$0.03&--\\
Sz 98               &K7$\pm$0.5     &4060$\pm$187   &1      &0.7$\pm$0.2  &3.2$\pm$0.7  &6.40$\pm$0.02&47$\pm$0\\
Sz 103              &M4$\pm$0.5     &3270$\pm$75    &0.7    &0.25$\pm$0.03  &1.4$\pm$0.3  &6.36$\pm$0.04&51$\pm$14\\
Sz 104              &M5$\pm$0.5     &3125$\pm$72    &0      &0.18$\pm$0.03  &1.1$\pm$0.3  &6.26$\pm$0.04&--\\
Sz 106              &M0.5$\pm$0.5   &3777$\pm$174   &1      &0.5$\pm$0.1  &0.7$\pm$0.2  &6.30$\pm$0.03&--\\
Sz 110              &M4$\pm$0.5     &3270$\pm$75    &0      &0.26$\pm$0.03  &1.6$\pm$0.4  &6.35$\pm$0.02&50$\pm$7\\
Sz 113              &M4.5$\pm$0.5   &3197$\pm$74    &1      &0.19$\pm$0.03  &0.8$\pm$0.2  &6.23$\pm$0.05&38$\pm$8\\
Sz 114              &M4.8$\pm$0.5   &3175$\pm$73    &0.3    &0.23$\pm$0.03  &1.8$\pm$0.4  &6.38$\pm$0.02&17$\pm$2\\
Sz 117              &M3.5$\pm$0.5   &3340$\pm$77    &0.50   &0.29$\pm$0.03  &2.0$\pm$0.4  &6.37$\pm$0.02&--\\
Sz 130              &M2$\pm$0.5     &3560$\pm$164   &0      &0.4$\pm$0.1  &1.0$\pm$0.2  &6.28$\pm$0.02&53$\pm$10\\
Sz 131              &M3$\pm$0.5     &4000$\pm$79    &1.9    &0.30$\pm$0.04  &1.0$\pm$0.2  &6.23$\pm$0.03&0$\pm$0\\
V856 Sco            &A7$\pm$0.5     &--             &--     &2.8$\pm$0.3  &--             &6.31$\pm$0.03&--\\
\enddata
\tablecomments{Spectral types, masses, and inclinations are from \citet{ansdell2016} and \citet{ansdell18}. Effective temperatures and radii are from \citet{alcala2017}. The extinction value of Sz 96 is from \citet{comeron2009}. The rest of the extinction values are from \citet{alcala2017}. Parallax data are from \citet{gaia2016,gaiadr3}. As noted in Section~\ref{analysis}, J16085373-3914367 was not included in our analysis.}
\end{deluxetable*}

\begin{figure*}[t!]     
\plotone{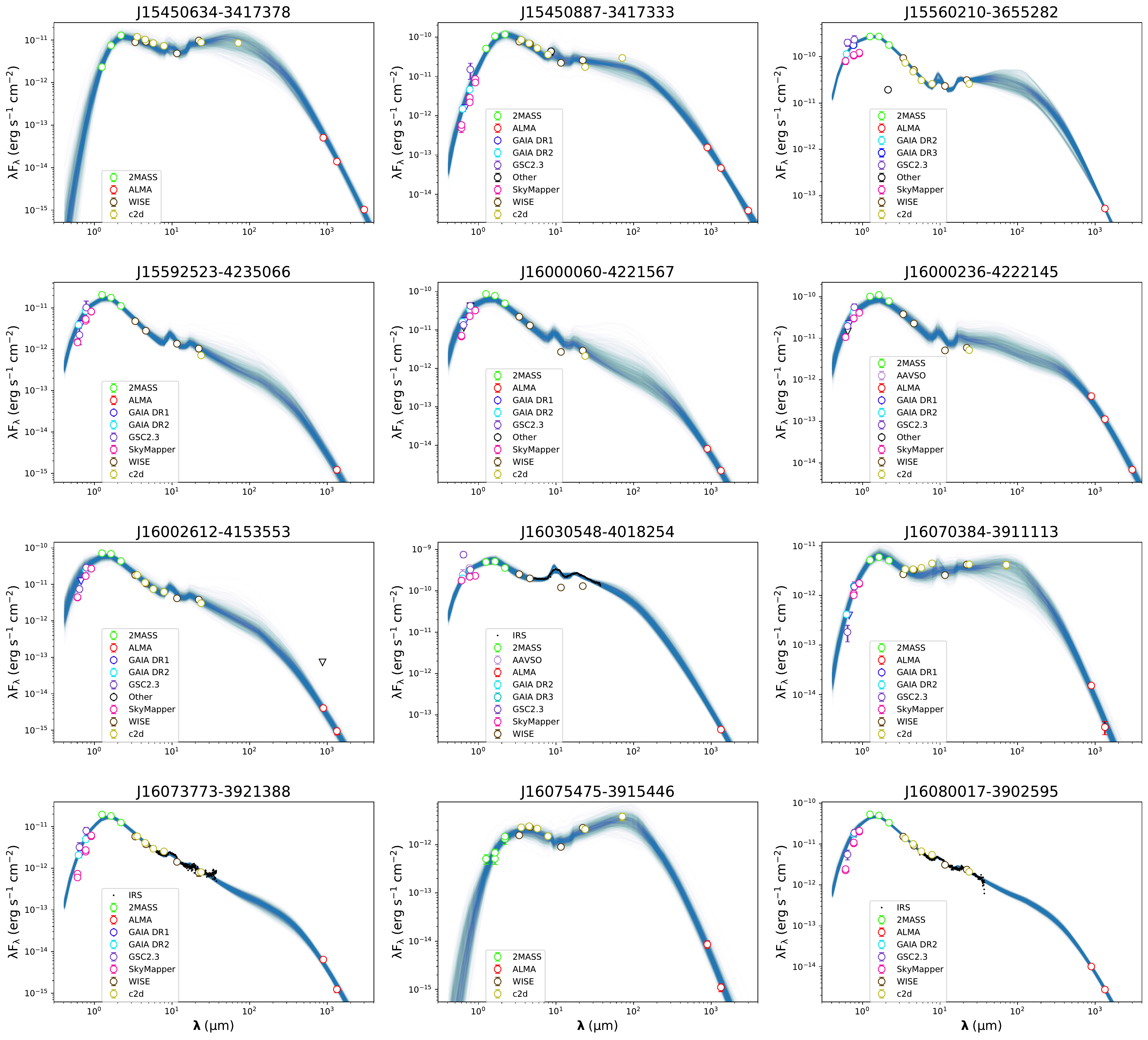}
\caption{
Posterior distributions (blue lines) of all sources that are modeled in this study. The models are reddened using the extinction law by \citet{mcclure2009} and extinction values obtained from the Markov chain Monte Carlo (MCMC) fitting process described in section \ref{sec:method} and plotted against the observed photometric (circles) and low-resolution spectroscopic (black dots) data.
}
\end{figure*} 

\begin{figure*}[t]     
\ContinuedFloat 
\plotone{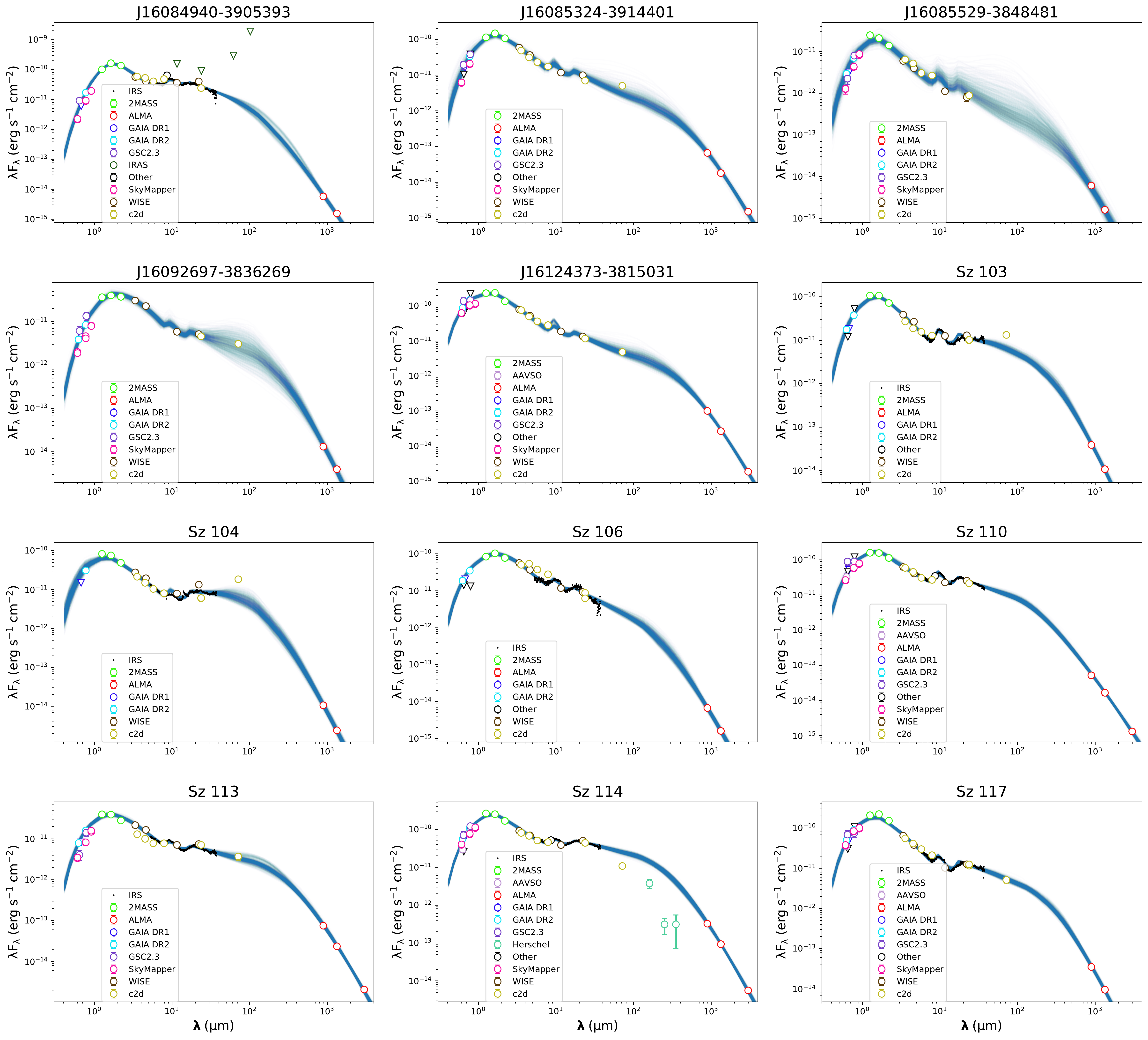}
\caption{
(continued)
}
\end{figure*} 

\begin{figure*}[t]     
\ContinuedFloat 
\plotone{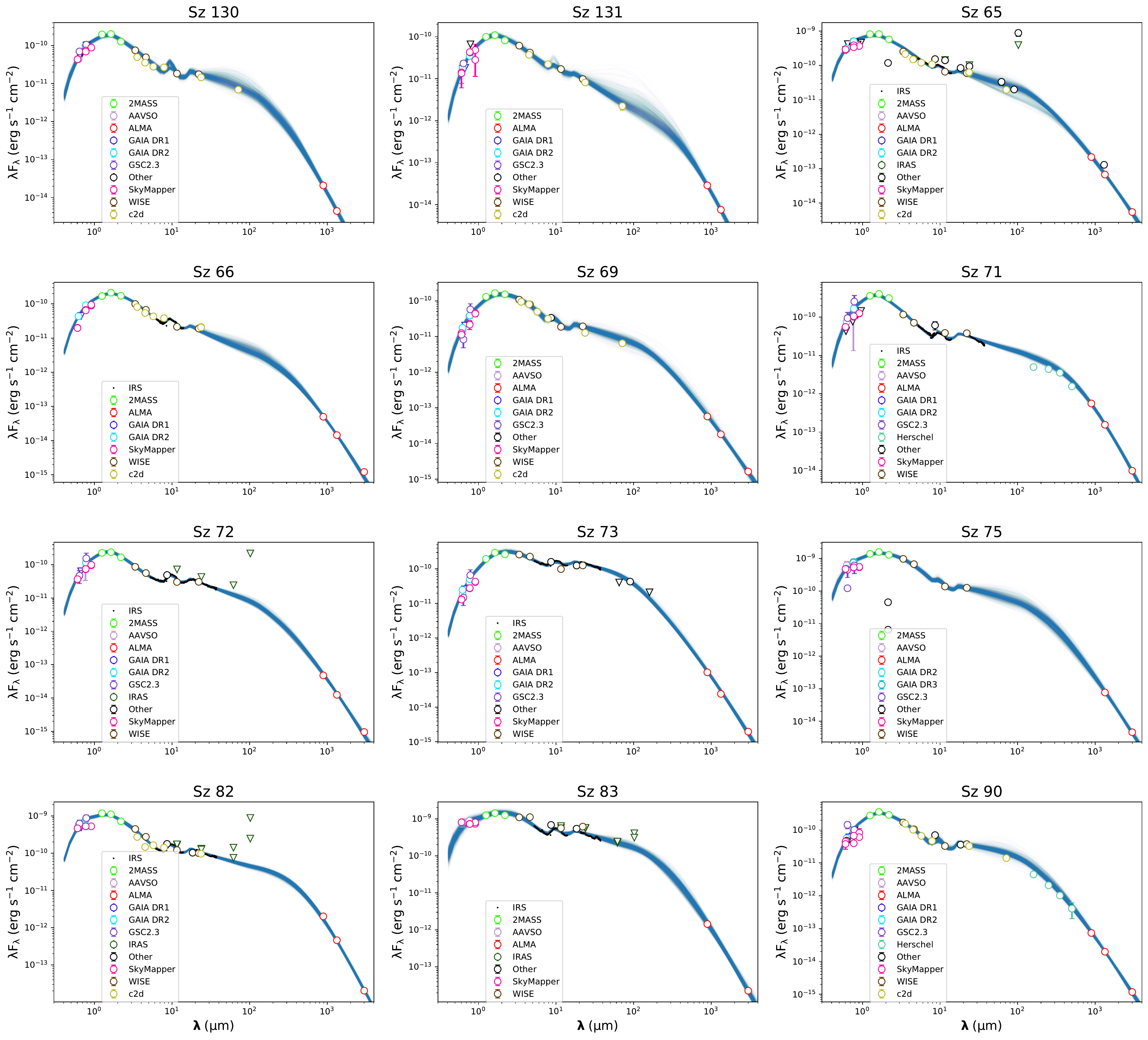}
\caption{(continued)}
\end{figure*} 

\begin{figure*}[t]     
\ContinuedFloat 
\plotone{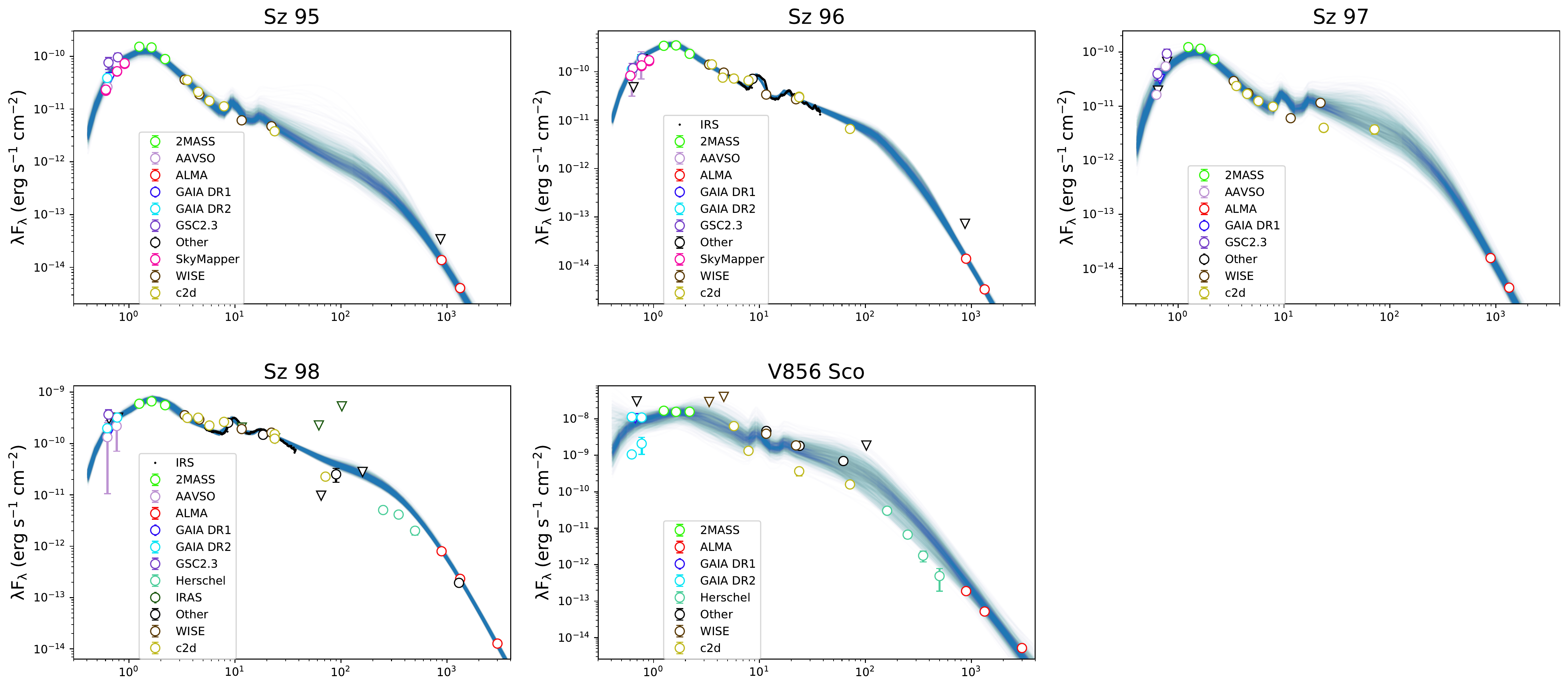}
\caption{
(continued)
}
\end{figure*} 

\begin{figure*}[t]      
\centering
\subfloat{\includegraphics[width=60mm]{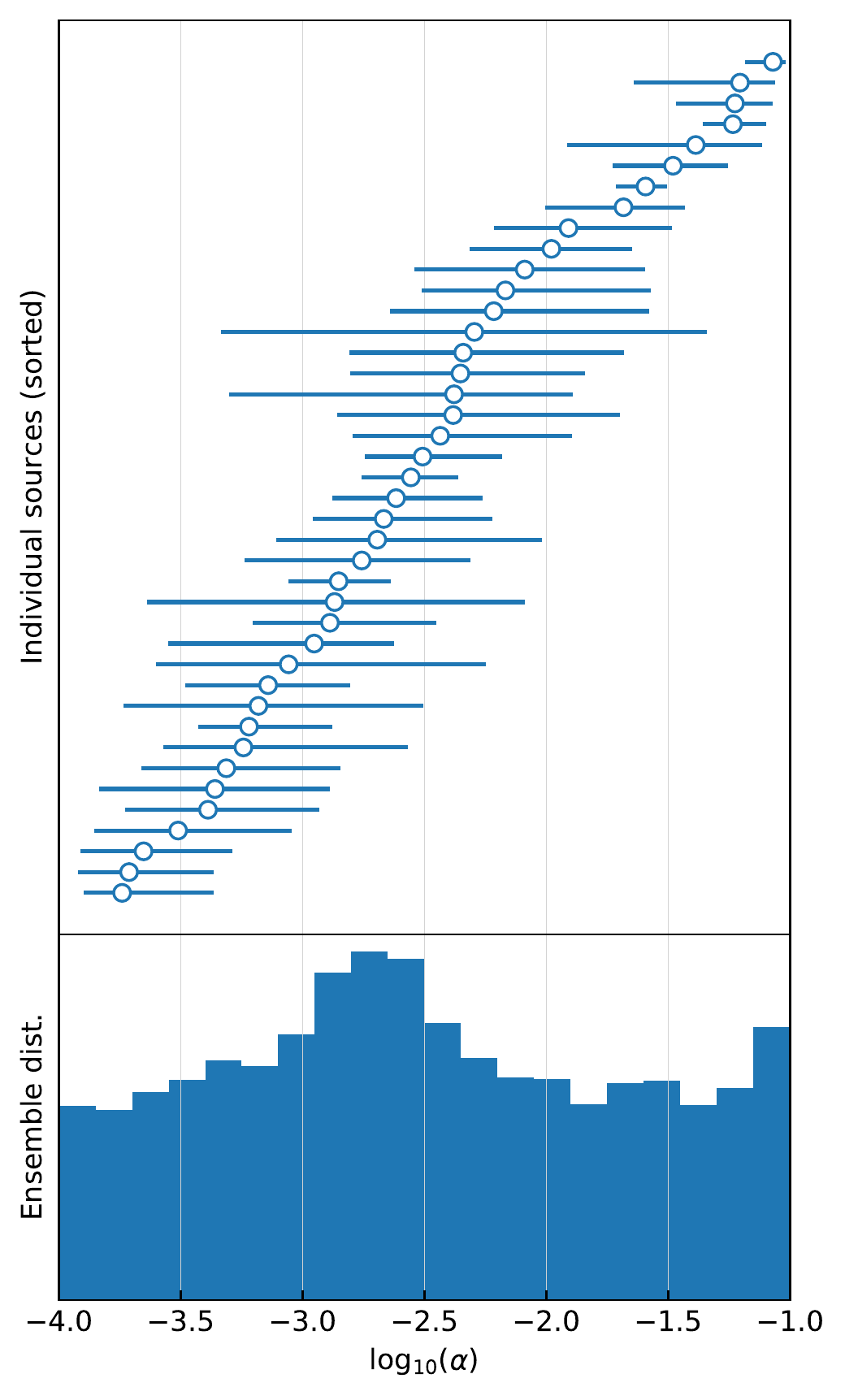}}
\subfloat{\includegraphics[width=60mm]{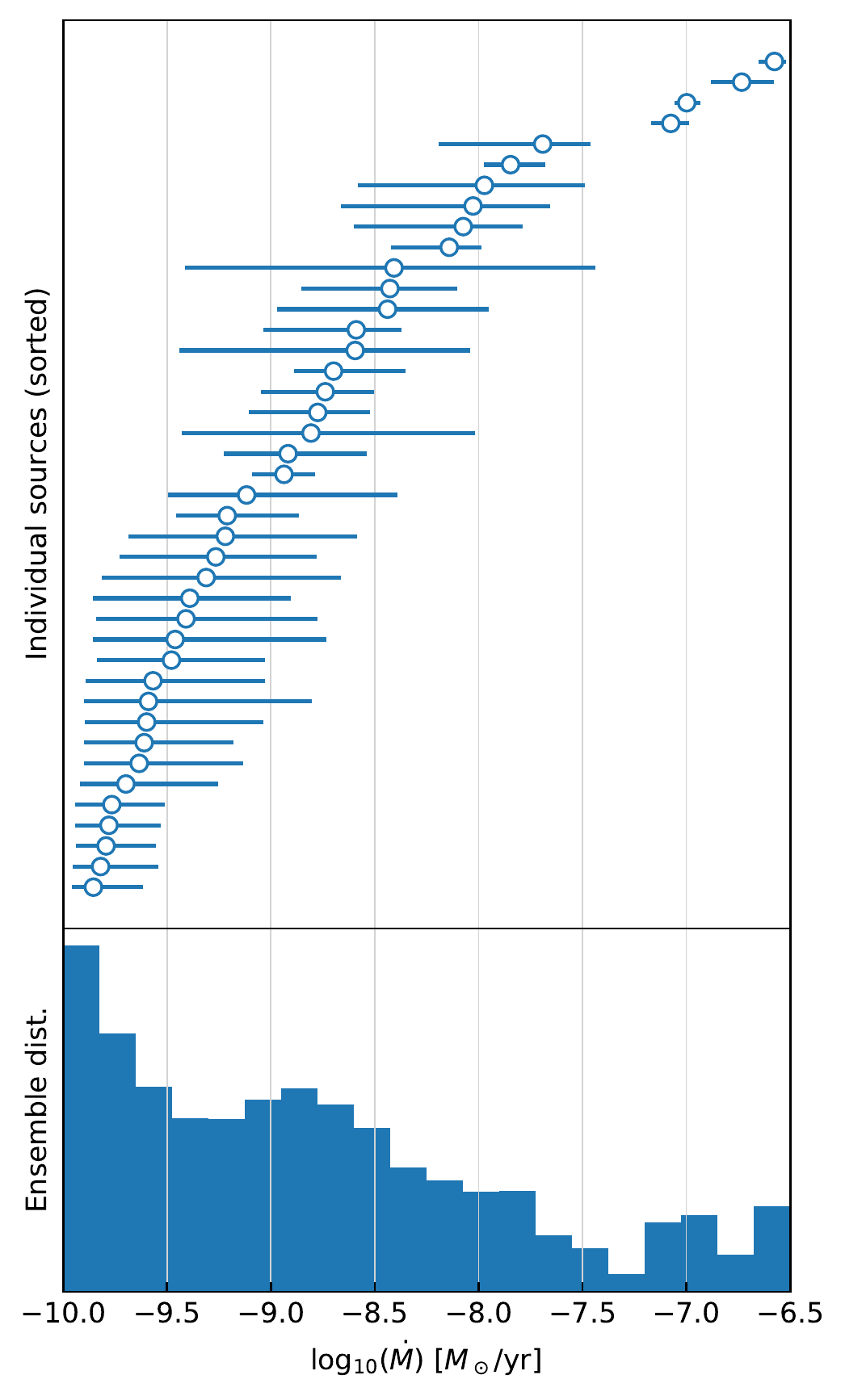}}
\subfloat{\includegraphics[width=60mm]{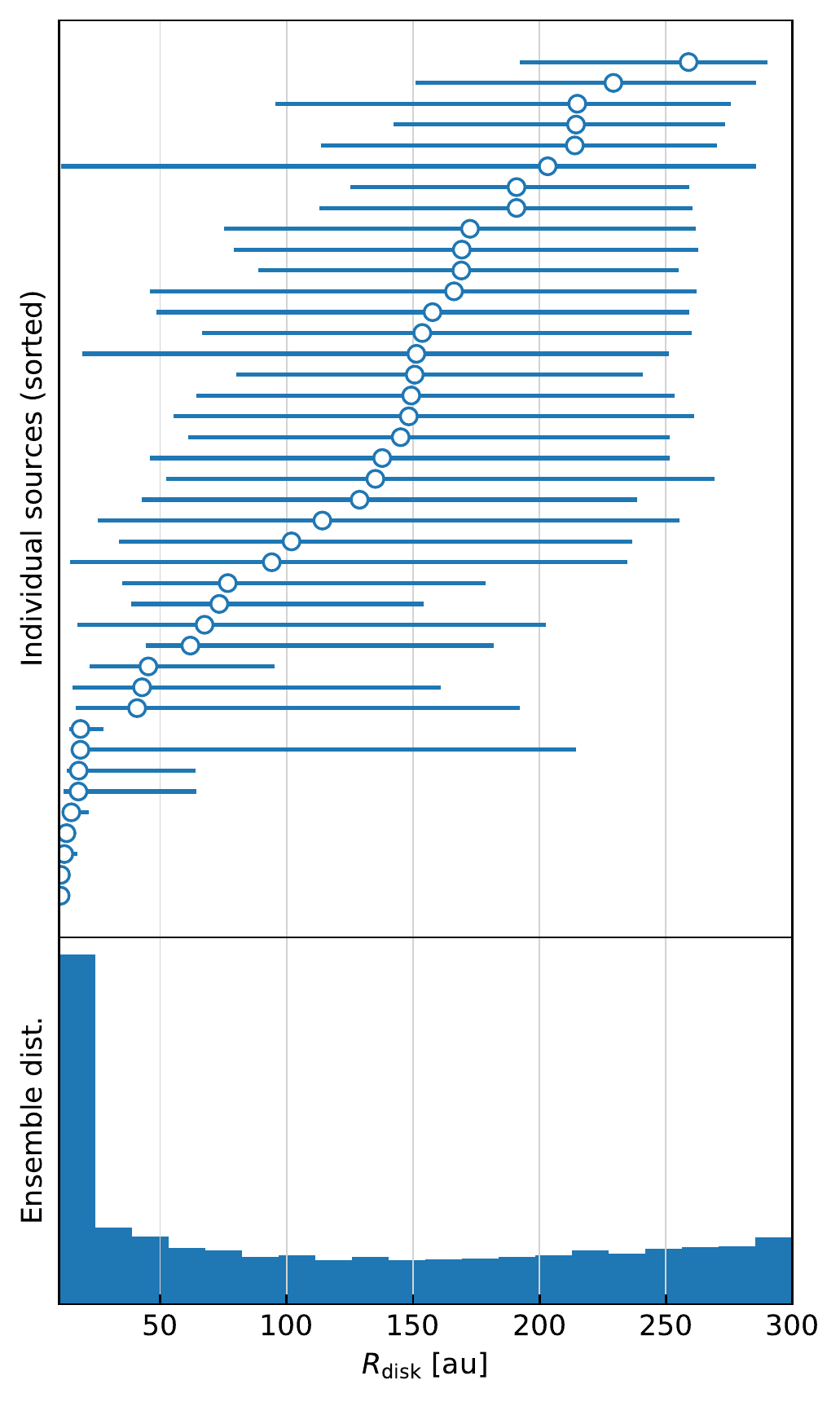}} \\
\subfloat{\includegraphics[width=60mm]{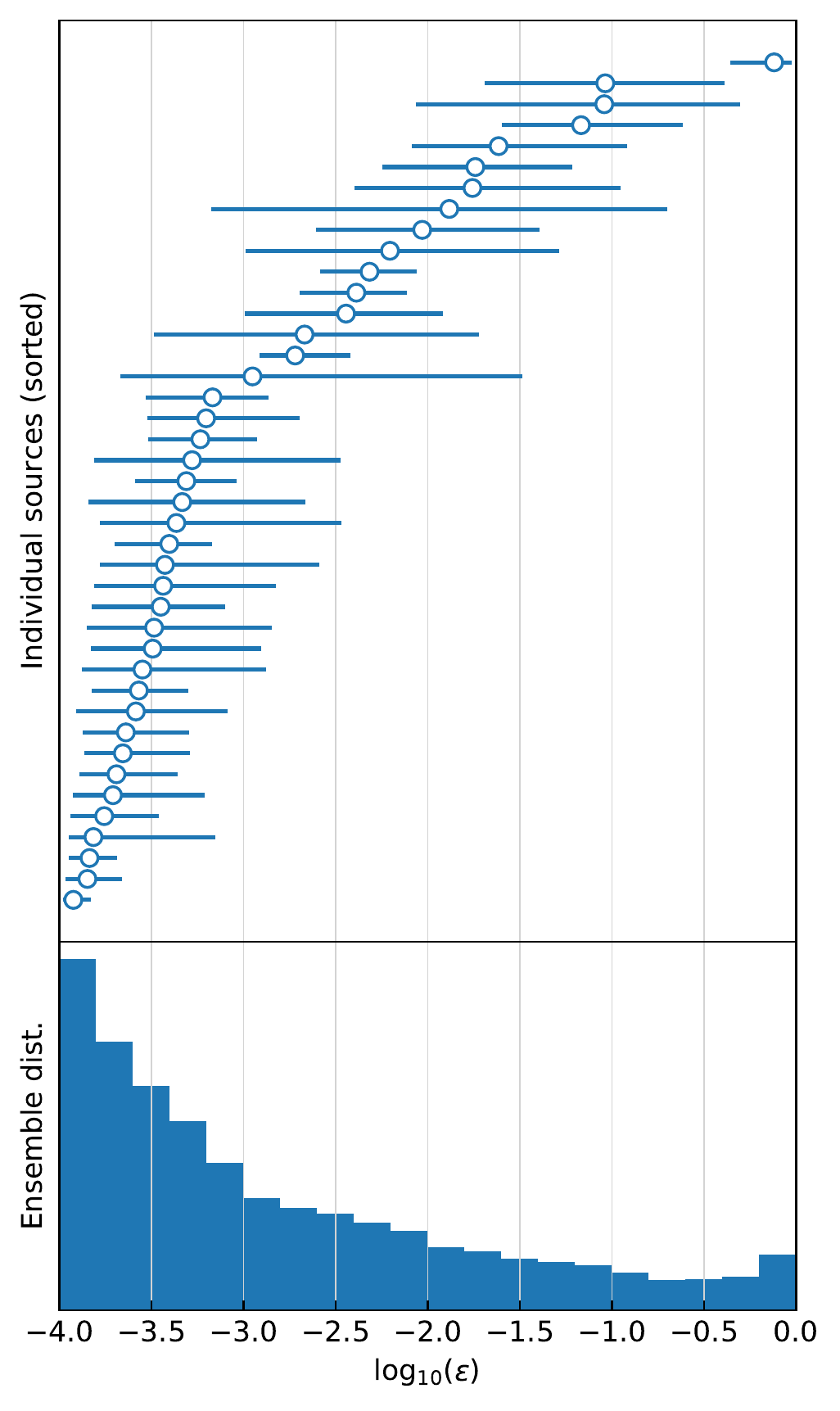}}
\subfloat{\includegraphics[width=60mm]{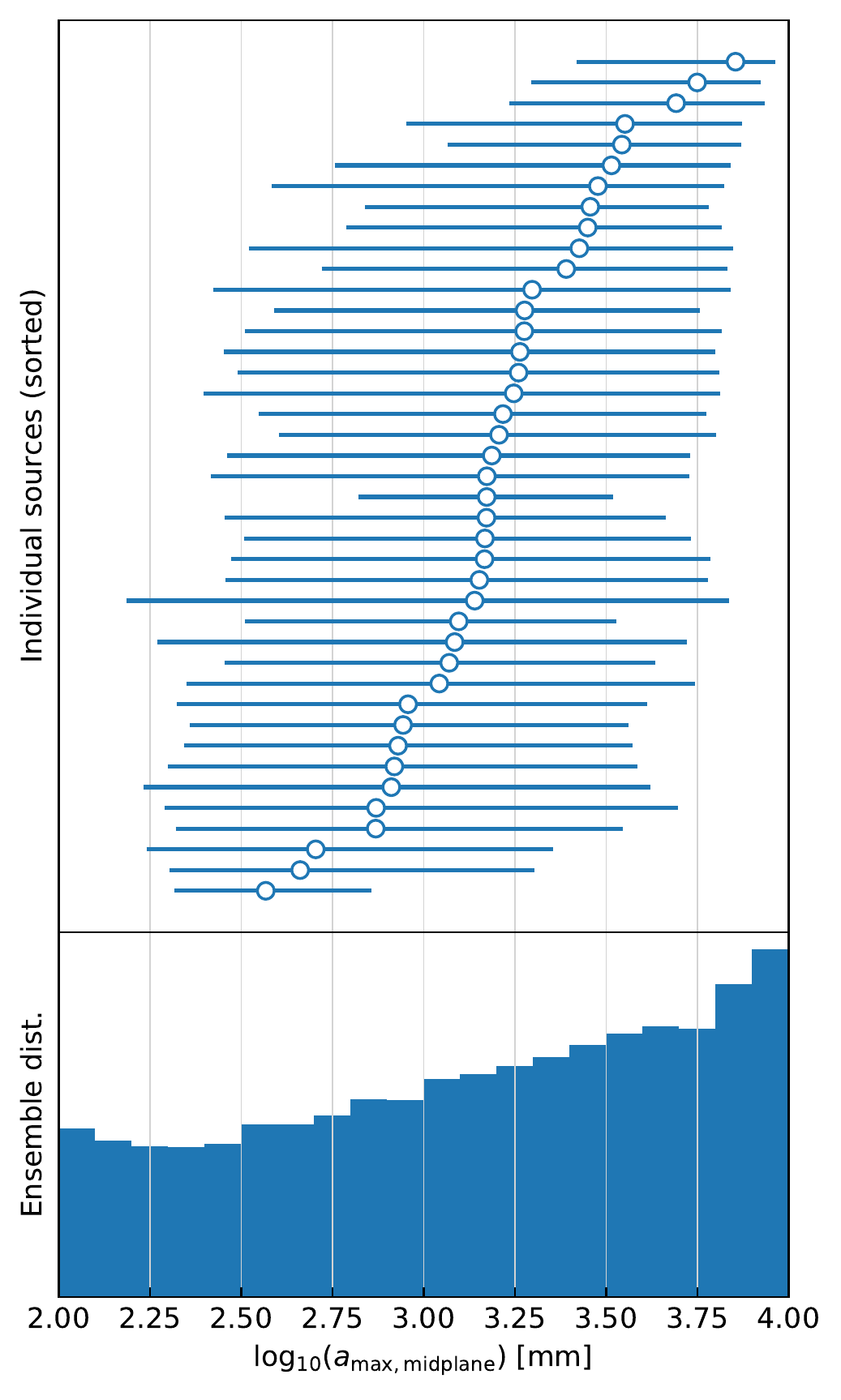}}
\subfloat{\includegraphics[width=60mm]{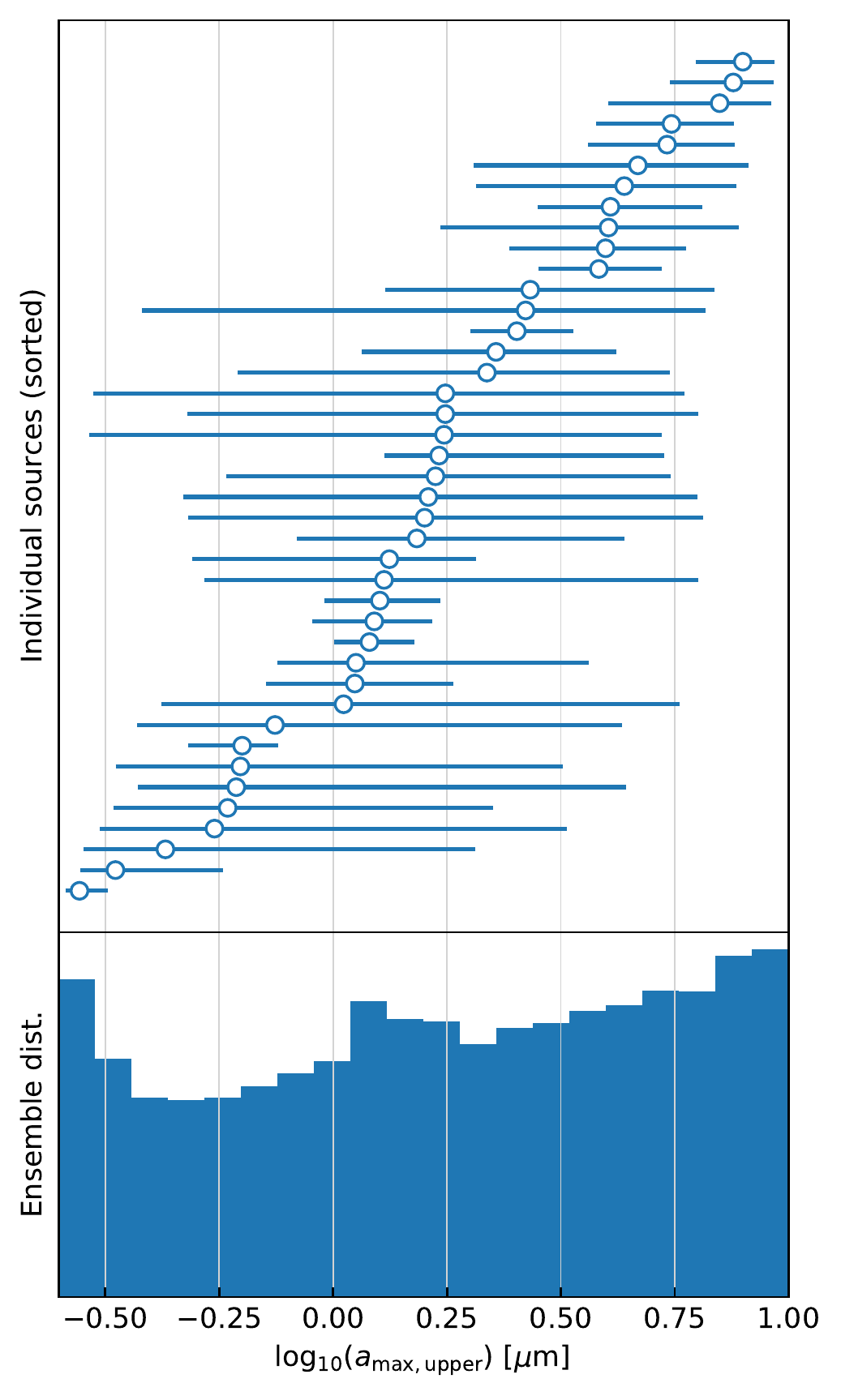}} \\

\caption{
Posterior distributions of individual sources (top panels) and the entire sample (bottom panels) for $\alpha$ (a), $\dot{M}$ (b), $R_{\rm disk}$ (c), $\epsilon$ (d), $a_{\rm max, midplane}$ (e), $a_{\rm max, upper}$ (f), M$_{\rm disk}$ (g), and M$_{\rm disk}/M_*$ (h). The error bars represent 16th and 84th percentiles. A total of 20 bins are used for the histograms in the bottom panels.
}
\end{figure*}      

\begin{figure*}[t]      
\ContinuedFloat 
\centering
\subfloat{\includegraphics[width=60mm]{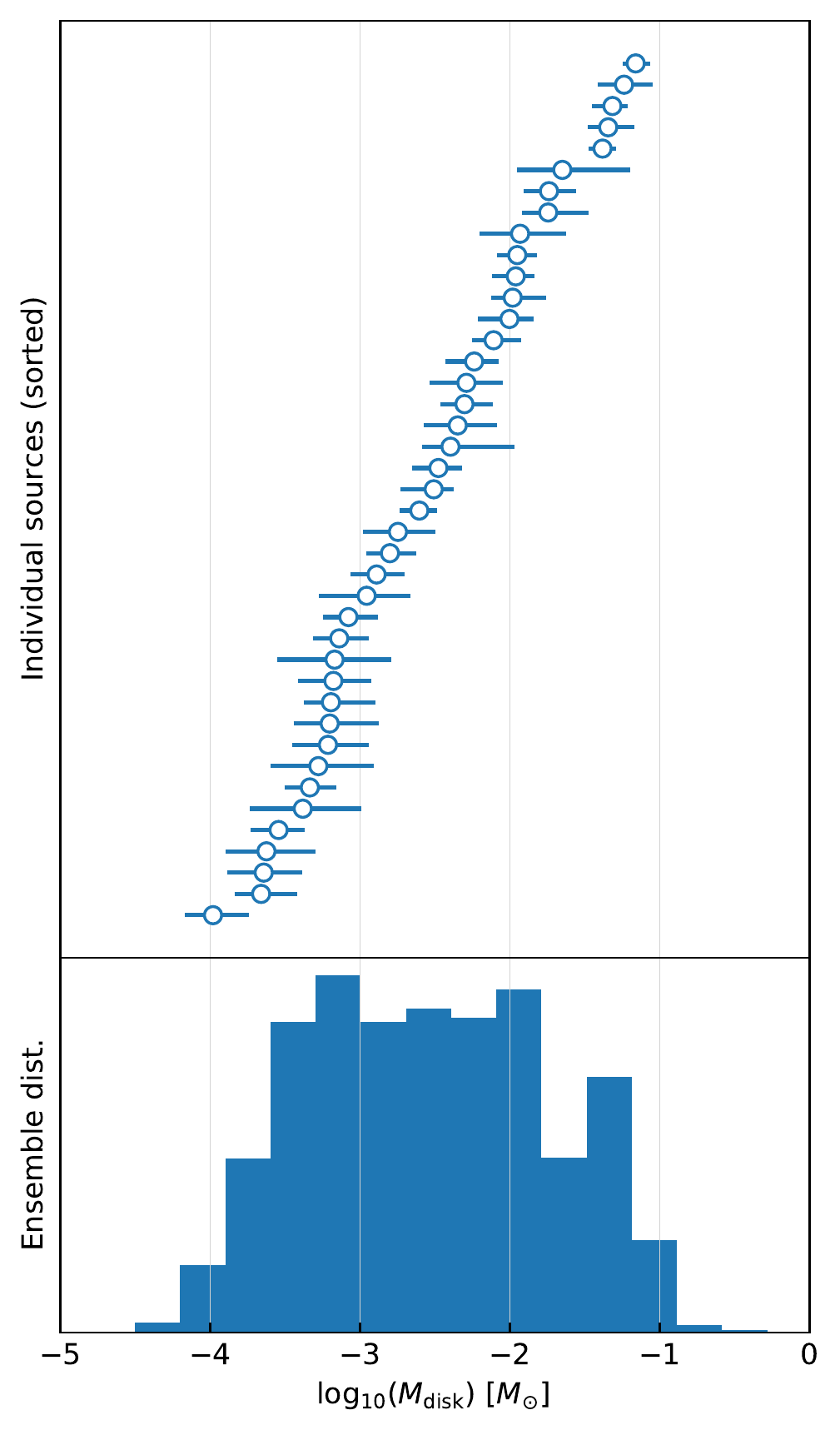}} \subfloat{\includegraphics[width=60mm]{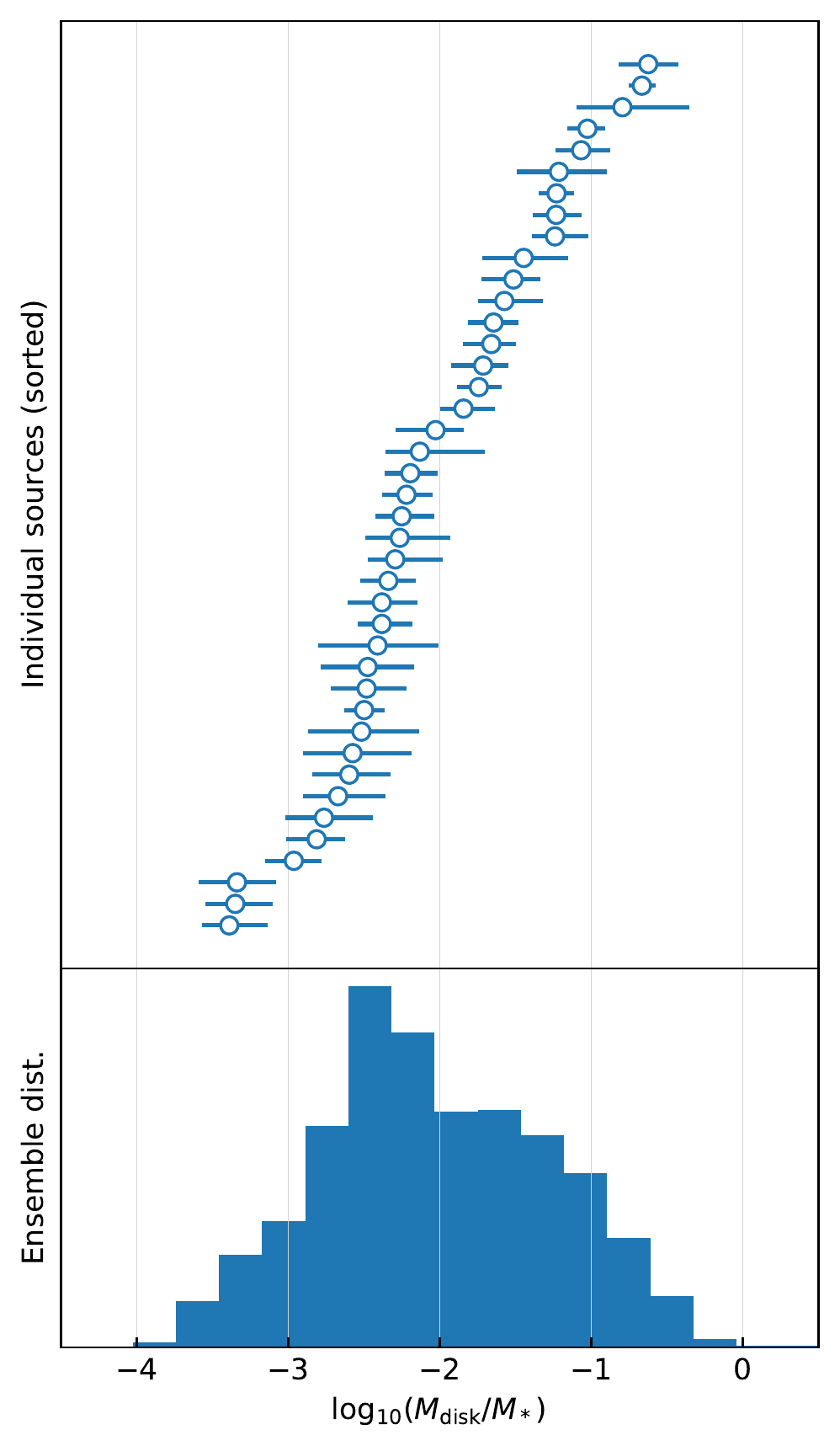}}\\
\caption{(continued)}
\end{figure*}      

(Sub-)millimeter data are critical in our fitting process since they constrain settling and $M_{\rm dust}$.  Therefore, we limited objects in our sample to those with existing (sub-)millimeter measurements from \citet{ansdell2016} and \citet{ansdell2018}.  
Of the 93 targets studied by \citet{ansdell2016} and \citet{ansdell2018}, 51 sources are discarded for the following reasons.
\begin{enumerate}
    \item Inner cavities of transitional disks (TDs) can have a substantial impact on their SEDs. Due to the complexity in modeling their radial structure, TDs are not included in our ANN. As a result, we discarded the sources that are identified as TDs in the literature. \citet{ansdell2016} identified 2MASS J16083070-3828268, RY Lup, and Sz~111 as TDs due to their clearly resolved inner cavities. In addition, the following sources are identified as TDs in the literature: MY Lup \citep{romero2012, vanmarel2018}, Sz~112 and 2MASS J16070854-3914075 \citep{ansdell2016, vanmarel2018}, 2MASS J16090141-3925119 \citep{marel22}, 2MASS J16011549-4152351 \citep{marel22}, 2MASS J16102955-3922144 \citep{bustamante2015, marel22}, 2MASS J16081497-3857145     \citep{bustamante2015, marel22}, Sz~84, Sz~100, Sz~118, and Sz~123A \citep{vanmarel2018}, and Sz~129 \citep{marel22}. After visually examining the compiled SEDs, 2MASS J16102741-3902299, and V* V1192 Sco have strong mid-infrared (MIR) dips that our model cannot fit, which may indicate these objects are TDs. We also remove these two objects from the sample. The aforementioned 17 sources are discarded from our sample.
    \item In order to constrain the SED models at radio wavelengths, we require that objects in our sample have (sub)millimeter flux densities. We discarded 24 sources that have flux uncertainties greater than 32\% (1-$\sigma$ level) for the ALMA photometry points.
    \item Due to the large number of free parameters in our model, a minimum number of MIR, JHK, and ALMA photometry points are required to achieve a reasonable fit. We require at least two JHK and MIR photometry points to inform the modeling of the stellar properties and trace the disk emission, respectively. 2MASS J16134410-3736462 has only one photometry point that is not an upper limit in the MIR region, and Sz~108b does not have 2MASS JHK photometry points available. Therefore, these two sources are excluded from our sample.
    \item Binaries and multiple systems could cause confusion when we compile their SED data and, in the case of low spatial resolution, have inaccurate photometry. Sz~88 A+B \citep{ansdell18}, Sz~68 (HT Lup A+B) and Sz~74 A+B \citep{tazzari2021} are binaries or components of multiple systems that were not resolved individually in their respective ALMA surveys. Near-infrared (NIR) adaptive optics imaging finds that these systems have separations of 245~au, 17~au, and 49~au, respectively \citep{zurlo21}. Sz~81A has a companion at a separation of 309~au \citep{ansdell18, zurlo21}. 2MASS J16095628-3859518 is a newly identified binary with a separation of $<90$~au \citep{zurlo21}. The aforementioned six sources are removed from our sample. Sz~66 is also a binary, but the separation is 983~au \citep{zurlo21}; the disk of the primary should not be affected by the secondary, given the large separation, and we leave it in our sample.
    \item Sz 102 (also known as Krautter's star) was excluded since it is known to be subluminous, falls below the Zero-Age Main Sequence, and hosts a strong outflow \citep{krautter86, hughes94}. Sz 133 was also removed from our sample since it is a subluminous object that falls below the Zero-Age Main Sequence \citep{alcala17}.
\end{enumerate}

For each of the sources remaining in the sample, we obtained optical to MIR photometry from several catalogues in VizieR \citep{Ochsenbein2000} as well as {\it Herschel} far-infrared (FIR) photometry and ALMA fluxes. See Table~1 for a complete list of the photometry used in this study. We also include low-resolution {\it Spitzer} Infrared Spectrograph (IRS) spectra from CASSIS \citep{Houck2004} when available.  Additionally, we compiled the stellar parameters of our sample from the literature and these are listed in Table~2.

\section{Methodology} \label{sec:method}
In this study, our goal is to estimate the posterior probability density function (posterior) for several disk parameters described further below. Therefore, we adopted a Bayesian framework for our analysis. The use of Bayesian analysis also allows us to incorporate preexisting information of the disk parameters from the literature.

\subsection{Models}

Due to their extensive computation time, the DIAD models are unrealistic to sample from when using a Bayesian approach. Thus, we used the set of ANNs developed by \citet{ribas2020}. Our ANN is trained on samples produced by DIAD and can mimic the effects of DIAD in a timescale magnitudes faster than that of DIAD. Using a set of stellar and disk parameters, DIAD computes the disk physical structure and its SED at a designated range of wavelengths. Due to the different nature of the DIAD outputs, our ANN contains two artificial neural networks: ANN$_{\rm SED}$ and ANN$_{\rm diskmass}$, which produce SED and disk mass estimates, respectively.

The disk parameters used by ANN are as follows:
\begin{itemize}
    \item the mass accretion rate in the disk, $\dot{M}$
    \item the disk viscosity, $\alpha$, following \citet{shakura73}.
    \item dust settling, $\epsilon$
    \item maximum grain sizes of the dust in the upper layer of the disk, $a_{\rm max, upper}$, and in the disk midplane, $a_{\rm max, midplane}$
    \item the disk radius, $R_{\rm disk}$
    \item the inclination of the disk, $i$ (important for ANN$_{\rm SED}$ only)
    \item the dust sublimation temperature, $T_{\rm wall}$
    \item the scaling of the disk inner wall, $z_{\rm wall}$.
\end{itemize}

Following \citet{ribas2020}, we use combinations of the stellar parameters $M_*$, $T_*$, and $R_*$ that are consistent with stellar evolution. We added a free variable, $Age_*$, which we used in conjunction with $M_*$ to calculate $T_*$ and $R_*$ based on the MESA Isochrones and Stellar Tracks. Then we use $M_*$ and the resulting consistent $T_*$ and $R_*$ pairs as input stellar parameters for ANN. The SED estimates obtained from ANN$_{\rm SED}$ are then reddened based on the interstellar extinction given by parameter $A_{\mathrm{v}}$, using the extinction law by \citet{mcclure2009}. Then, the reddened SEDs are scaled to the correct distance calculated from the parallax parameter.

In order to take into account possible systematic uncertainties that may be present in the data from different instruments, we scaled all of the uncertainties in the flux densities according to a free parameter, $f$. Some photometry points may differ significantly from the rest of the SEDs and are problematic in the fitting process. Thus, we employed a mixture model as a means to combat the effects of possible outliers in our sample \citep{hogg2010}. The three parameters used in the outliers model are: $P_{\rm out}$, the probability that any photometry point is an outlier, and $y_{\rm out}$ and $V_{\rm out}$, the mean and variance of the outliers.

Hence, a total of 17 free parameters are used in the fitting process: the nine disk parameters, $Age_*$, $M_*$, $A_{\mathrm{v}}$, parallax of the source, the three parameters used in the mixture model, and $f$. The SEDs and models are shown in Figure~1 and discussed further in Section~4.

\subsection{Priors}

For most of the disk parameters, we use flat priors, with the exception of $T_{\rm wall}$. A temperature of 1400~K is typically assumed for the dust sublimation temperature, and thus we use a Gaussian centered at 1400~K and a standard deviation of 50~K for $T_{\rm wall}$. We use Gaussian priors for $M_*$, parallax, and inclination listed in Table~2 to represent preexisting information on these parameters (using their uncertainties as the standard deviation) and flat priors when no literature values are available. All priors used for our ANN's input parameters have bounds set by the parameter space of our ANN's training set. We refer the reader to Appendix~B.2 of \citet{ribas2020} for the bounds of ANN's training set. Priors for $A_{\rm v}$ and parallax have limits set to reasonable ranges for Lupus: $A_{\rm v}$ is set between 0 and 8, and parallax is calculated to be between 5~mas and 8.33~mas (corresponding to distances between 120~pc and 200~pc).

\subsection{Likelihood Functions}

We incorporated information from four data sets in the fitting process: {\it Spitzer} IRS spectral data, photometry data, $T_{\rm eff}$, and $R_*$. We used Gaussian likelihood functions for IRS spectral data, $T_{\rm eff}$, and $R_*$. The photometry data are divided into two categories: ``general'' and ``critical'' photometry. 2MASS and ALMA photometry data do not overlap other photometry data in the wavelength ranges that they occupy and are crucial for constraining several of our parameters. To avoid misclassification as outliers by the mixture model outlined in Section~3.1, we included photometry data from 2MASS and ALMA in the critical photometry data set and the rest of the photometry in the general photometry data set. We employed a standard Gaussian likelihood function for the critical photometry data set and the mixture model for the general photometry data set. The IRS spectra have substantially more points than the rest of the SED and thus are binned at the closest output wavelength of ANN using the method of weighted mean.

As described in Section~3.2, we are scaling all of the uncertainties in the fluxes according to the parameter $f$. 

\begin{equation}
    s_n^2 = \sigma_{n,{\rm model}}^2 +\sigma_{n,{\rm obs}}^2 +f^2\,y_{n,{\rm obs}}^2
\end{equation}

\noindent where index $n$ corresponds to each measurement, $\sigma_{n,{\rm model}}$ is the adopted 10\% uncertainty for our ANN's prediction at the same wavelength, $\sigma_{n,{\rm obs}}$ is the uncertainty in the flux for the measurement, and $y_{n,{\rm obs}}$ represents the observed fluxes of the measurement.

The Gaussian likelihood function used for spectral data and critical photometry data is given by

\begin{equation}
\text{$\mathcal{L}$}_{n,{\rm spect}}= \frac{1}{\sqrt{2\pi (s_n^2)}} \exp{\biggl(-\frac{(y_{n,{\rm spect}} - y_{n,{\rm model}})^2}{2(s_n^2)}\biggr)}
\end{equation}

\begin{equation}
\text{$\mathcal{L}$}_{n,{\rm crit}}= \frac{1}{\sqrt{2\pi (s_n^2)}} \exp{\biggl(-\frac{(y_{n,{\rm crit}} - y_{n,{\rm model}})^2}{2(s_n^2)}\biggr)}
\end{equation}

\noindent where index $n$ corresponds to each measurement, and $y_{n,{\rm spect}}$ and $y_{n,{\rm crit}}$ represent the observed fluxes of the spectral data and 2MASS/ALMA photometry, respectively, $y_{n,{\rm model}}$ represents the fluxes predicted by our ANN at the same wavelength.

The outliers mixture model that we adopted from \citet{hogg2010} to model the general photometry data has a likelihood function given by

\begin{dmath}
\text{$\mathcal{L}$}_{n,{\rm gen}}= \frac{1-P_{\rm out}}{\sqrt{2\pi(s_n^2)}} \exp{\biggl(-\frac{(y_{n,{\rm gen}} - y_{n,{\rm model}})^2}{2(s_n^2)}\biggr)}
 + \frac{P_{\rm out}}{\sqrt{2\pi [s_n^2 + V_{\rm out}^2]}} \exp{\biggl(-\frac{(y_{n,{\rm gen}} - y_{\rm out})^2}{2 [s_n^2 +
    V_{\rm out}^2]}\biggr)}
\end{dmath}

\noindent where $y_{n,{\rm gen}}$ represents the observed fluxes of the photometry points.

The Gaussian likelihood function used for $T_{\rm eff}$ and $R_*$ are given by

\begin{equation}
\text{$\mathcal{L}$}_{T_{\rm eff}}= \frac{1}{\sqrt{2\pi (\sigma_{T_{\rm eff}}^2)}} \exp{\biggl(-\frac{(T_{\rm eff} - T_{\rm eff,{\rm model}})^2}{2(\sigma_{T_{\rm eff}}^2)}\biggr)}
\end{equation}

\begin{equation}
\text{$\mathcal{L}$}_{R_*}= \frac{1}{\sqrt{2\pi (\sigma_{R_*}^2)}} \exp{\biggl(
-\frac{(R_* - R_{*,{\rm model}})^2}{2(\sigma_{R_*}^2)}\biggr)}
\end{equation}

\noindent where $T_{\rm eff}$, and $R_*$ are the values of effective temperature and radius of the star reported in the literature, and $T_{\rm eff,model}$ and $R_{*,{\rm model}}$ are the values of effective temperature and radius of the star produced by the MESA isochrones. We note that adopting a Gaussian likelihood for $R_*$ could introduce inconsistencies in the modeling process if the literature $R_*$ values are obtained using a different isochrone or pre-{\it Gaia} distance estimates. However, in our testing, we found no significant differences between using a Gaussian likelihood function for $R_*$ and using a flat likelihood function for $R_*$.

The overall likelihood function takes into account the information from spectral data and photometry. Additionally, due to the use of an Markov chain Monte Carlo (MCMC) sampler, we use the likelihood functions and priors in log space. Thus, the overall likelihood function becomes

\begin{dmath}
\ln\mathcal{L} = \sum_{n=1}^{N_{\rm gen}}\ln\mathcal{L}_{n,{\rm gen}} + \sum_{n=1}^{N_{\rm crit}}\ln\mathcal{L}_{n,{\rm crit}} 
+ \sum_{n=1}^{N_{\rm spect}}\ln\mathcal{L}_{n,{\rm spect}} + \ln\text{$\mathcal{L}$}_{T_{\rm eff}} + \ln\text{$\mathcal{L}$}_{R_*}~~.
\end{dmath}

\subsection{MCMC Method}

In this work, we used the parallel tempering implementation of Goodman \& Weare’s Affine Invariant MCMC Ensemble sampler, $\it{ptemcee}$, to approximate the potentially multimodal posterior. We use three temperatures and the minimum number of walkers required by $\it{ptemcee}$ to speed up performance. We ran for 200,000 steps in total to fully sample the posterior. Visual inspection of the chain evolution indicates that convergence occurs before 50,000 steps for all sources in our sample, and thus we discarded the first 100,000 steps of the chains and use only the last 100,000 steps to obtain the posteriors that are used in the subsequent analysis.

\begin{figure*}[ht!]     
\plotone{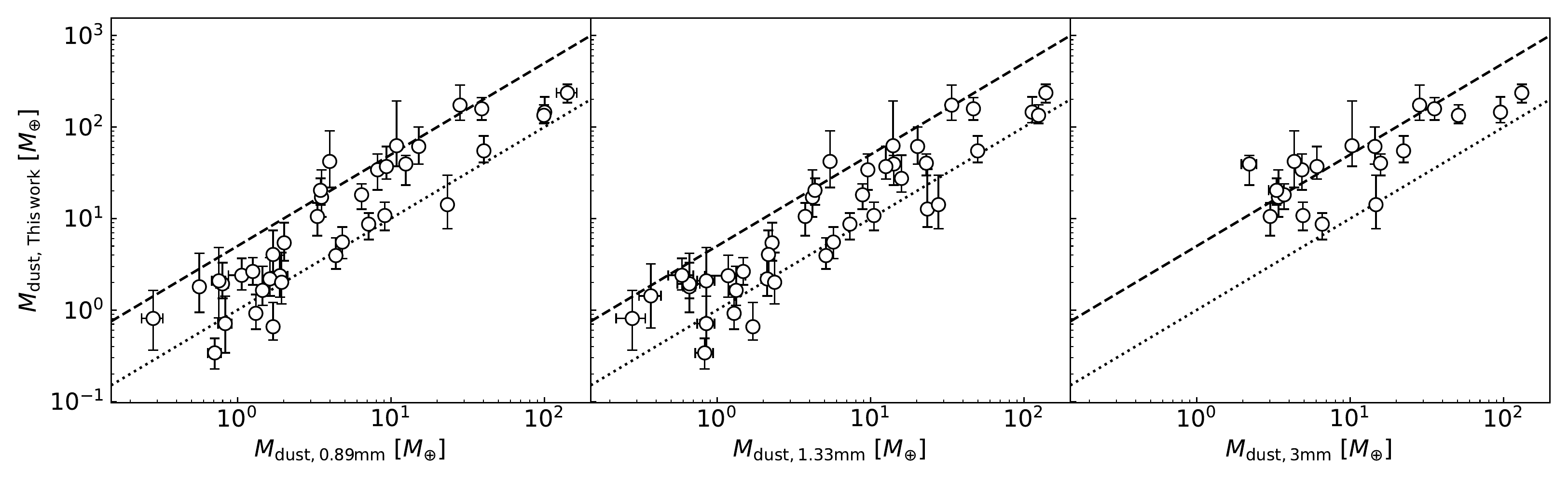}
\caption{
Disk dust masses produced by this work using SED modeling and disk dust masses reported in the literature using 0.89~mm fluxes \citep[left;][]{ansdell2016},  1.33~mm fluxes \citep[middle;][]{ansdell2018}, and 3~mm fluxes \citep[right;][]{tazzari2021}. The one-to-one relation is plotted as a dotted line and a five-to-one relation is plotted as a dashed line. Note that only the sources that are covered by the 3~mm ALMA survey are plotted in the right panel \citep{tazzari2021}. Error bars from our work represent 16th and 84th percentiles of the samples.\label{fig:compare}
}
\end{figure*} 

\section{Analysis and Results} \label{analysis}

In order to visualize the SED models associated with the posterior of each source, we randomly selected 1000 samples from the posterior and plotted them against the photometry and IRS spectra. The SEDs and models are shown in Figure~1. Bimodality occurs for multiple objects with IRS spectra in our sample (e.g., J16084940-3905393, Sz 82, Sz 83, Sz 98). This is to be expected since the numerous data points from IRS spectra greatly constrain the fits in the MIR region, yet disagreement between data points from {\it Herschel} and ALMA can skew the fits in the FIR, creating multiple modes in the region. Bimodality could cause large error bounds in the posterior distributions of the model distributions. We note that the fit to J16085373-3914367 is very poor and so we remove it from the rest of the analysis. The corner plots for the MCMC fits of individual sources are available on Zenodo\footnote{\url{https://doi.org/10.5281/zenodo.7251263}}.

In order to characterize any physical trends in our sample, we performed marginalization for parameters of interest for our 41 sources. We plot the marginalized distribution for each relevant parameter as error bars for individual sources and a histogram for the entire sample. The resulting graphs are shown in Figure~2 for $\alpha$ (a), $\dot{M}$ (b), $R_{\rm disk}$ (c), $\epsilon$ (d), $a_{\rm max, midplane}$ (e), $a_{\rm max, upper}$ (f), M$_{\rm disk}$ (g), and M$_{\rm disk}/M_*$ (h).

The disk mass is not a free parameter and depends on the parameters $\alpha$ and $\dot{M}$, which determine the surface density profile, $\Sigma$ $\propto$ $\dot{M}$/$\alpha$ (see Eqn. 37 in D'Alessio et al. 1998). Larger values of $\alpha$ therefore correspond to smaller disk masses for a given $\dot{M}$, and vice versa. These two parameters are generally correlated, but the individual effects of $\dot{M}$ and $\alpha$ on an SED can help break the degeneracy between these parameters. Tests performed by \citet{ribas2020} indicate that $\alpha$ mostly affects SED estimates in the MIR and FIR range, while $\dot{M}$ has a significant impact in the wavelengths of the IRS spectra due to the heating by the accretion luminosity \citep{ribas2020}. As shown in the corner plots for each object, either $\dot{M}$, $\alpha$, or both are well-constrained for each object, which is sufficient to constrain the disk mass.

We note that there are nine objects in our sample with observed accretion rates below 10$^{-10}$ $M_{\odot}$ yr$^{-1}$ \citep{alcala2017}, which is below the range for which the ANN was trained. Given the variable nature of accretion (e.g., Fischer et al. 2022) and that chromospheric activitybf, which depends on stellar mass, sets a lower limit on robust measurements of mass accretion rates below $\sim10^{-9}$ $M_{\odot}$ yr$^{-1}$ for young ($\sim$1 Myr), solar mass objects \citep{manara13}, we keep these nine objects in our sample. We note that in our SED modeling $\dot{M}$ is the mass accretion rate in the disk, while the above $\dot{M}$ from the literature is the mass accretion rate measured onto the star; these two $\dot{M}$ are not necessarily the same.

$R_{\rm disk}$ (Figure~2c) appears to be unconstrained for most of our sources, often spanning most of the parameter space. This parameter appears to have no distinctive effect on the overall SED estimates \citep{ribas2020}, suggesting that SED fitting alone cannot fully constrain $R_{\rm disk}$. Note that, even if the disk radius is not constrained, that does not prevent the disk mass from being constrained, as the disk mass is also determined by other parameters  that affect the SED.

The marginalized distribution of $\epsilon$ (Figure~2d) shows a preference toward the lower end, indicating overall high levels of dust settling for our sample. This finding is consistent with previous work using SED modeling that has found that dust is depleted by up to 1000 times relative to the interstellar medium \citep{ribas2020, grant18} with the same levels of high depletion found in regions across the 1--10 Myr age range (Rilinger et al. 2022, submitted). This evidence of significant dust settling supports that planet formation may may be well underway by 1~Myr.

$a_{\rm max, midplane}$ (Figure~2e) is unconstrained for most of our sources and has no clear impact on the simulated SED \citep{ribas2020}. 
Some sources have well constrained $a_{\rm max, upper}$ (Figure~2f) while others do not. The discrepancy is likely to be caused by the availability of IRS spectra, which is unsurprising since $a_{\rm max, upper}$ most strongly affects the same range of wavelengths that the IRS spectra occupy. Our results show the importance of IRS spectra in constraining the dust population characteristics in the disk atmosphere. However, we note that we did not explore various dust compositions, which have important effects in the shape of the silicate feature.

Our SED modeling produced disk mass estimates that vary by order of magnitudes for the entire sample. The 16th, 50th, and 84th percentiles of M$_{\rm disk}$ (Figure~2g) are 0.0004, 0.003, 0.2 {\msun} and for M$_{\rm disk}/M_*$ (Figure~2h) are 0.002, 0.008, 0.07 {\msun}. We discuss our measured disk dust masses further in Section~5.

\begin{figure*}[ht!]     
\plotone{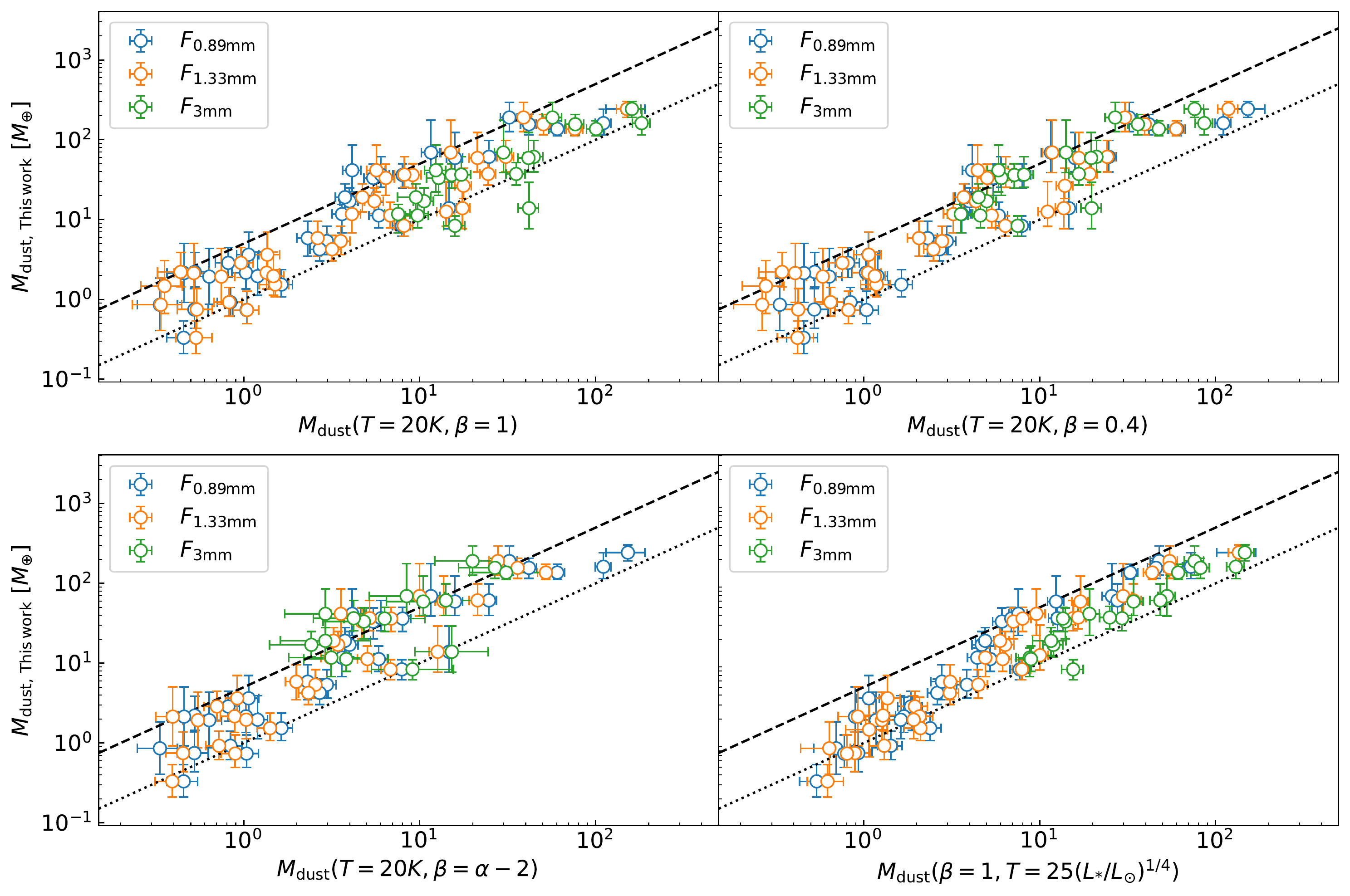}
\caption{
Comparison of the disk dust masses produced by this work using SED modeling with disk dust masses obtained from the \citet{hildebrand83} dust mass equation using different sets of assumptions. We use parallax data from Gaia EDR3 to calculate the distances used in the dust mass equation. The first and second panels assume a constant $\beta$ value of 1 and 0.4, respectively, and a constant dust temperature of $T=20$K. The third panel uses a constant dust temperature of $T=20$K and $\beta=\alpha-2$, where $\alpha$ is the spectral index calculated from the fluxes at 0.89~mm and 1.33~mm for each object. The fourth panel uses a constant $\beta=1$ and the host star luminosity scaling from \citet{andrews13}, $T=25(L_{*}/L_{\odot})^{1/4}$~K, where $L_{*}$ is the luminosity of the central source for each object. 0.89~mm fluxes (blue points) are from \citet{ansdell2016}, 1.33~mm fluxes (orange points) are from \citet{ansdell2018}, and 3~mm fluxes (green points) are from \citet{tazzari2021}. The one-to-one relation is plotted as a dotted line and a five-to-one relation is plotted as a dashed line. Note that only the sources that are covered by the 3~mm ALMA survey are plotted in green \citep{tazzari2021}\label{fig:betas}.
}
\end{figure*} 

\section{Discussion}

In the following sections, we discuss our SED modeling measurements of the disk dust masses of 41 disks in Lupus in comparison to dust masses previously reported in the literature.  We then consider that the differences between our masses and those previously reported in the literature are due to the disk's optical depth at millimeter wavelengths.  We explore the effect of disk optical depth on the measured dust masses, utilizing the fact that Lupus has the largest sample of disks in one region that has been observed to date at 3~mm with ALMA at high sensitivity \citep{tazzari2021}.

\subsection{Comparison to Disk Dust Masses in the Literature}

We compared the measurements of disk dust mass produced by our work and in the literature and find that our masses are generally higher. In Figure \ref{fig:compare}, we plot the disk dust masses produced by this work compared with the disk dust masses produced by \citet{ansdell2016} using 0.89~mm fluxes, by \citet{ansdell18} using 1.33~mm fluxes, and by \citet{tazzari2021} using 3~mm fluxes. The aforementioned works use the common assumption that the disk is isothermal and optically thin at (sub)millimeter wavelengths and thus that the disk dust mass is proportional to the (sub)millimeter flux following \citet{hildebrand83}:
$$M_{\rm dust} = \frac{F_\nu d^2}{\kappa_\nu B_\nu(T_{\rm dust})}$$\label{eq:dustmass}
\noindent where $F_\nu$ is the flux at observed frequency $\nu$, $d$ is the distance to the object, $\kappa_\nu$ is the dust opacity at frequency $\nu$, and $B_\nu(T_{\rm dust})$ is the Planck function for a characteristic dust temperature of $T_{\rm dust}$. However, different dust opacities and dust temperatures are used in the literature. In \citet{ansdell2016} and \citet{ansdell2018}, a dust grain opacity, $\kappa_\nu$, of 10~cm$^2$~g$^{-1}$ at 1000~GHz is adopted using a power-law index of $\beta=1$ following \citet{beckwith90}. \citet{tazzari2021} adopted a dust grain opacity $\kappa_\nu=3.37(\nu/337~{\rm GHz})^{\beta}$~cm$^2$~g$^{-1}$, where $\beta$ is related to the spectral index $\alpha$ as $\beta=\alpha-2$. All three works adopted a constant dust temperature of $T_{\rm dust}=20$K, the median dust temperature of disks located in Taurus-Auriga derived by \citet{andrews18}.  

To facilitate a more direct comparison between our disk masses and those inferred from (sub)millimeter fluxes, we recalculated the millimeter-derived disk masses using commonly assumed values of $\beta$ and temperature and plotted them against our SED-derived disk masses (which include the 3~mm point in the fit when available) in Figure \ref{fig:betas}. We find that our SED-derived disk masses are still higher than those calculated using the \citet{hildebrand83} equation. 
The millimeter-derived disk masses using the \citet{andrews13} temperature scaling relation and $\beta=1$ (Figure 4d) yield the best agreement with our models. \citet{andrews13} proposes that $T_{\rm dust}$ scales weakly with $L_*$ due to thermal radiation from the central source heating the regions where most of the millimeter-wave emission is generated and adopts the scaling $T_{\rm dust}=25(L_*/L_\odot)^{1/4}K$. 

The SED-derived disk masses are 1.5--6 times higher than the millimeter-derived disk masses. The median ratios between the SED-derived disk masses and the millimeter-derived disk masses for the entire sample at all ALMA wavelengths are 1.5--5.4 for Figures 4 a, b, and c, and 1.5--2.1 in Figure~4d.
As also seen in Rilinger et al.\ (2022, submitted), the discrepancy between the SED-derived disk masses and the millimeter-derived disk masses is even larger at the higher disk masses. We calculate the median ratios between the SED-derived disk masses and the millimeter-derived disk masses at each ALMA wavelength using a cutoff point of 10 M$_{\earth}$. In Figure~4d, the ratios are 3.5, 2.7, and 1.6 at 0.89mm, 1.3mm, and 3mm, for the subset of sources that have disk masses $>$10~M$_{\earth}$.  For the subset of sources that have disk masses $<$10~M$_{\earth}$, the ratios are 1.2, 1.2, and 0.6, respectively. A larger discrepancy between between the SED-derived disk masses and the millimeter-derived disk masses at higher disk masses is also seen in Figures 4 a, b, and c where the ratios are 3.7, 2.7--5.0, 1.6--5.8 for $>$10~M$_{\earth}$ and 2.0, 1.5-2.0, 0.5-1.2 for $<$10~M$_{\earth}$.

\subsection{The Effect of Disk Optical Depth on Measured Dust Masses}

\citet{ballering19}, \citet{ribas2020}, \citet{liu22}, and Rilinger et~al.\ (2022, submitted) found that disk masses derived from SED modeling are higher than those derived from millimeter flux densities alone, and they attributed this to optical depth effects. Masses derived from a single millimeter flux density measurement depend on the assumption that the disk is optically thin at that wavelength; if the disk is optically thick, this assumption results in an underestimate of the disk mass because the measurement is not sensitive to all of the dust. 
We note that disk masses derived from our SED modeling are not subject to the same assumption. The emission from the disk at a given wavelength basically depends on the optical depth along the line of sight and the temperature structure of the disk. If the disk is optically thick, the emission will be given mainly by the disk temperature at the $\tau_{\lambda}=1$ surface. DIAD computes the radial and vertical disk density and temperature structure using the $\alpha$-disk prescription, heating the disk with viscosity and stellar irradiation, and enforcing hydrostatic equilibrium.  As such, the disk structure is calculated self-consistently. Therefore, DIAD can constrain many disk parameters even when a disk is optically thick at most wavelengths.

\begin{figure*}[ht!]     
\plotone{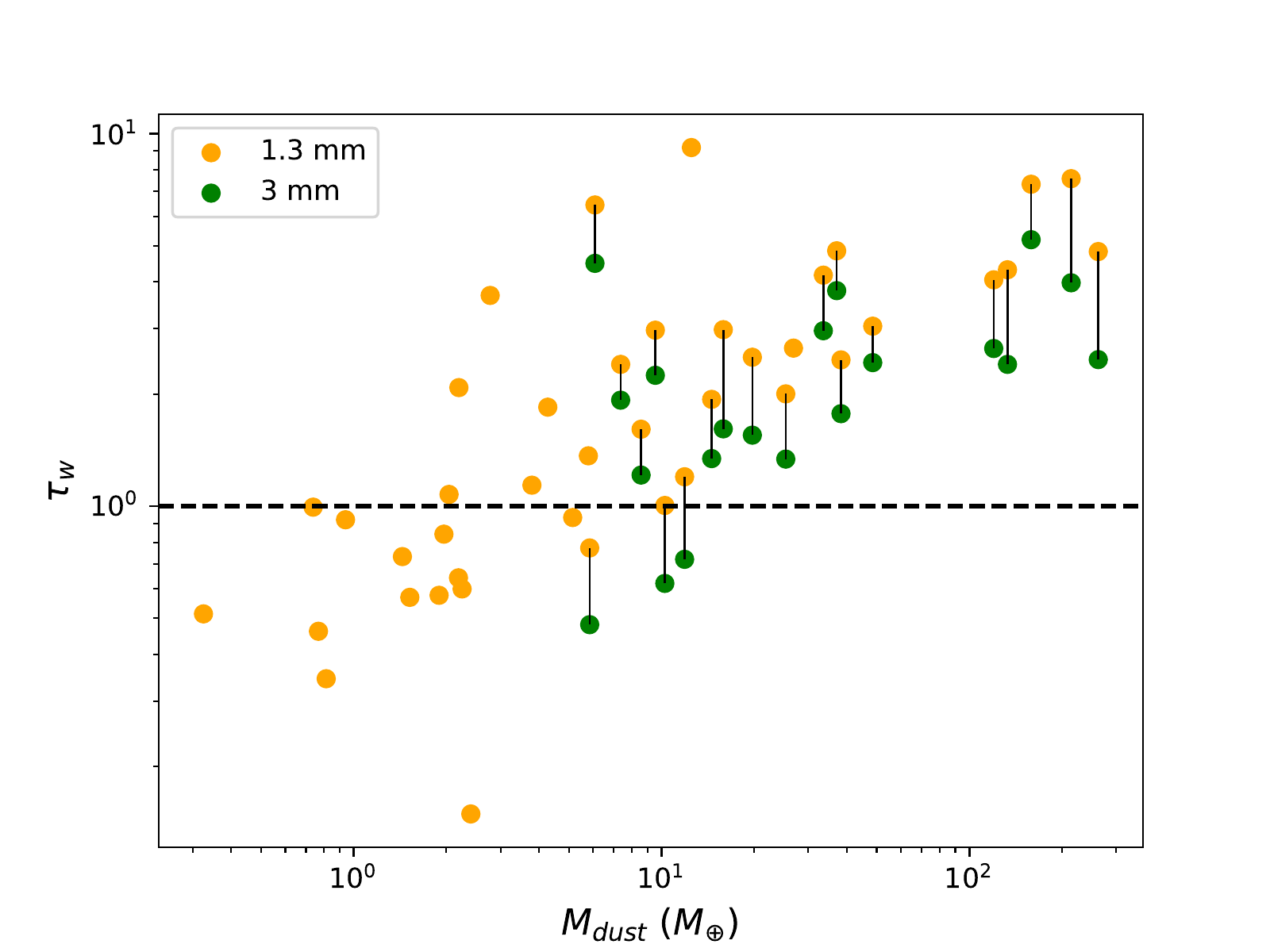}\label{fig:optdepth}
\caption{Flux-weighted mean optical depth of each disk plotted versus disk mass. Optical depths are flux-weighted averages of the optical depth at each radius in the disk as calculated by the DIAD models. Disk masses are the median values from our SED models for each object. The horizontal dashed line denotes an optical depth of 1. Orange points show the optical depth at 1.3 mm for each object, and green points represent the optical depth at 3 mm for the objects that have 3 mm flux measurements. Black vertical lines connect points for the same object.}
\end{figure*} 

To further study the disk optical depth in the millimeter, we plot the flux-weighted mean optical depth for each disk in our Lupus sample versus the disk masses obtained via our SED modeling in Figure~5. To determine the optical depths of the disks at millimeter wavelengths, we ran DIAD models for the disks in our sample. We used the median values of each model parameter for each object as determined by our SED fitting. DIAD calculates the disk structure and optical depth along the line of sight as a function of radius at the desired wavelength. Using the modeled flux at each radius as the weight, we calculated the weighted mean of the optical depths at each radius. This flux-weighted mean optical depth takes into account the disk regions which produce most of the emission at the given wavelength. We used a wavelength of 1.3 mm for all of the objects in our sample, and repeated the process at 3 mm for the 20 objects in our sample with 3 mm flux measurements from \citet{tazzari2021}.  

We find optical depths to be larger for more massive disks, and many of the optical depths to be close to or greater than 1 at both 1.3~mm and 3~mm (Figure~5). This shows that, as expected, the discrepancy between the millimeter-derived dust masses and those obtained from SED modeling is larger for the more massive disks due to their higher optical depths. Additionally, the optical depth at 3 mm is consistently lower than the optical depth at 1.3 mm, but is still greater than 1 for most objects. The optically thin assumption thus does not hold for most of the sources observed at 3 mm in our sample. \citet{macias21} first suggested that the 3mm emission may still be partially optically thick in analysis of TW~Hya. Our results in Lupus are the first confirmation of this using a sample of several disks with 3mm data.

Overall, these results showcase an important challenge currently faced by demographic studies aiming at measuring dust masses from optically thin mm fluxes: most of the disks that are optically thick at 1.3 mm are still so at 3 mm. In our modeling, the optical depth between these two wavelengths decreases by a factor of $\sim$1.5--2, which means that only disks with optical depths between 1 and 2 at 1.3 mm might be optically thin at 3 mm. While the $\sim$50$\%$ less massive disks in our sample are optically thin at 3 mm, observations at longer wavelengths are required to trace optically thin emission in the more massive disks. The Very Large Array and the upcoming ALMA Band 1 will be key for this.

Nevertheless, the slight decrease in optical depth at 3 mm could still help obtain more accurate masses. For example, \citet{ueda22} finds a range of disk masses, depending on the millimeter wavelength used. Their dust mass measured at 0.89~mm is $\sim$4 times smaller than that at 3.56~mm \citep{ricci10}. This is consistent with the disk being optically thinner at longer wavelengths, so more of the disk material is detected at longer wavelengths.

To further explore the effect of the disk optical depth at 3~mm on the measured disk masses, we fit the SEDs of the 20 disks in our sample that were detected at 3~mm with ALMA \citep{tazzari2021}, both including and excluding the 3~mm data from the SED in the fit. We find that the disk dust masses derived with our SED modeling when including the 3~mm data in the SED are on average 60--70\% higher than those disk dust masses derived when fitting the SED without the 3~mm ALMA data. This supports that the disk is optically thinner at 3~mm compared to 1.3 mm, as indicated in Figure ~5. If the disk had the same optical thickness at both wavelengths, the derived masses would be similar regardless of whether we included the 3~mm data in the SED fitting. The derived masses are higher when including the 3~mm data in the SED fitting since the disk is optically thinner at 3~mm and so, as noted earlier, the measured emission at this wavelength is tracing more (but still not all) of the dust that is present in the disk. There is also less discrepancy between the millimeter-derived and SED-modeling-derived disk dust masses at 3~mm, especially when using $\beta=1$ and the \citet{andrews13} temperature relationship (Figure~4d, green points). However, there is not total agreement, supporting the results in Figure~5 that the disk can be optically thick even out to 3~mm. 

While we note that SED modeling finds higher disk masses than calculations using millimeter flux densities alone, this is still likely an underestimate of the disk dust mass. First, here we are measuring only millimeter and smaller grains. Larger grains in the disk may exist that we are not sensitive to. It is also likely that significant substructure may be present in our sample \citep[e.g.,][]{jennings22}, and our model assumes no substructure.  Such substructures may be optically thick in the millimeter and ``hide'' even more mass \citep{tripathi17, andrews18}. Although, recent work by \citet{liu22} suggests that the mass hidden in optically thick substructures would not not significantly influence the measured dust mass. Many works have found that the mass in solids in exoplanetary systems is similar to that in protoplanetary disks \citep{najita14, manara18, mulders21}, suggesting a nearly 100\% planet formation efficiency \citep{mulders21} and hence that planet formation has to be earlier. Our results help to alleviate this tension and, when considered together with the high levels of dust settling in our sample, suggest that the conditions are right for planets to form as early as 1 Myr, and that they could still form within the protoplanetary disk phase.

\section{Summary}

In this study, we compiled the SEDs of protoplanetary disks in the Lupus star-forming region. Then, using a Bayesian framework and DIAD models, we modeled 41 protoplanetary disks and constrained their disk properties. Our main results are:

\begin{itemize}
    \item SED-modeling-derived disk mass estimates are generally $\sim1.5$--6 times higher than those calculated using (sub)millimeter flux densities alone. This discrepancy is caused by the high optical depths of the disks at (sub-)mm wavelengths. Our modeling shows that most of the disks in Lupus are optically thick at 1.3 mm.   
    \item The optical depth at 3 mm decreases only by a factor 1.5-2. As a consequence, most disks observed at this wavelength are still optically thick, and their dust masses are still underestimated. Nonetheless, the slightly lower optical depths can provide more accurate dust masses.
    \item Dust masses derived using the \citet{hildebrand83} disk dust mass equation along with the \citet{andrews13} temperature scaling relation ($T=25(L_{*}/L_{\odot})^{1/4}$~K) and $\beta=1$ are in best agreement with dust mass estimates measured with SED models, especially at 3~mm.
\end{itemize}

Future disk surveys should be undertaken with ALMA Band 1 (6–-8.6~mm) in order to trace optically thin emission and obtain more accurate disk dust masses.

\acknowledgements
We thank the referee for constructive feedback. This paper utilizes the D’Alessio irradiated accretion disk (DIAD) code. We wish to recognize the work of Paola D’Alessio, who passed away in 2013. Her legacy and pioneering work live on through her substantial contributions to the field.  We thank A. M. Hughes for insightful discussion. This work was funded by NASA ADAP 80NSSC20K0451. We acknowledge support from the NRAO Student Observing Support Program through award SOSPA8-007. A.R. has been supported by the UK Science and Technology research Council (STFC) via the consolidated grant ST/S000623/1 and by the European Union’s Horizon 2020 research and innovation programme under the Marie Sklodowska-Curie grant agreement No. 823823 (RISE DUSTBUSTERS project). This work has made use of data from the European Space Agency (ESA) mission {\it Gaia} (\url{https://www.cosmos.esa.int/gaia}), processed by the Gaia Data Processing and Analysis Consortium (DPAC, \url{https://www.cosmos.esa.int/web/gaia/dpac/consortium}). Funding for the DPAC has been provided by national institutions, in particular the institutions participating in the Gaia Multilateral Agreement.

\bibliographystyle{aasjournal}
\bibliography{bib_xin.bib}
\end{document}